# IOTA Proton Injector Beamline Installation[*]

D. Edstrom Jr[†], D. Broemmelsiek, K. Carlson, J.-P. Carneiro, H. Piekarz, A. Romanov, A. Shemyakin, A. Valishev, Fermi National Accelerator Laboratory, Batavia, IL, United States

*Abstract*

The IOTA Proton Injector (IPI), currently under installation at the Fermilab Accelerator Science and Technology facility, is a beamline capable of delivering 20-mA pulses of protons at 2.5 MeV to the Integrable Optics Test Accelerator (IOTA) ring. First beam in the IPI beamline is anticipated in 2023, when it will operate alongside the existing electron injector beamline to facilitate further fundamental physics research and continued development of novel accelerator technologies in the IOTA ring. This report details the expected operational profile, known challenges, and the current state of installation.

## INTRODUCTION

A Proton source and beamline is being installed at the Fermilab Accelerator Science & Technology (FAST) facility. This IOTA Proton Injector (IPI) will provide protons to the Integrable Optics Test Accelerator (IOTA) ring at 2.5 MeV to allow for studies on non-linear beam dynamics, electron lens research, electron cooling, and other novel accelerator technologies as Stage III of the R&D proposal that has driven development of the FAST facility [1]. The IPI is divided into four sections: the ion source, the low-energy beam transport (LEBT), the radio-frequency quadrupole (RFQ), and the medium-energy beam transport (MEBT). A selection of source, beamline, and IOTA ring parameters are summarized in Table 1.

## PROTON SOURCE

Protons are to be generated with kinetic energy of 50 keV using a Duoplasmatron ion source (IS) configured for proton production as described in *figure 1*. This source was used as a part of the High Intensity Neutrino Source (HINS) experiment [2], and was subsequently adopted by the FAST facility as it fits the desired beam parameters for proton production in the IPI beamline, as shown in *figure 2*. It was resurrected briefly in place in the HINS cave at the Meson Detector Building (MDB) before being fully disassembled and cleaned. In the process of disassembly, it was noted that the extraction aperture of 1.25 mm in use with HINS was unusually large for a duoplasmatron source, so this has been replaced with a 0.7 mm aperture. It has since been reassembled at the FAST facility.

Filaments are prepared on-site using a mixture of Barium, Calcium, and Strontium Carbonates along with Isoamyl Acitate to assist in binding the mixture to the filament. This is dissolved in Acetone to form a thin solution and the cleaned, Nickel filament is dipped to deposit a modest coating of the carbonate mixture [3]. It is then activated in a filament activation station referred to simply as the Jar (Fig. 3) by slowly increasing the filament current to the nominal 21 A, as seen in Figure 4. This burns away the nitrocellulose binder and ultimately converts the carbonate mixture into oxides [4], lowering the work function and facilitating electron transmission on the filament surface. This is checked at points by applying a bias voltage between the filament and a plate connected to ground. We then flash the filament by raising the filament supply current to 22 A for short periods and check the work function by monitoring the current draw from the HV supply.

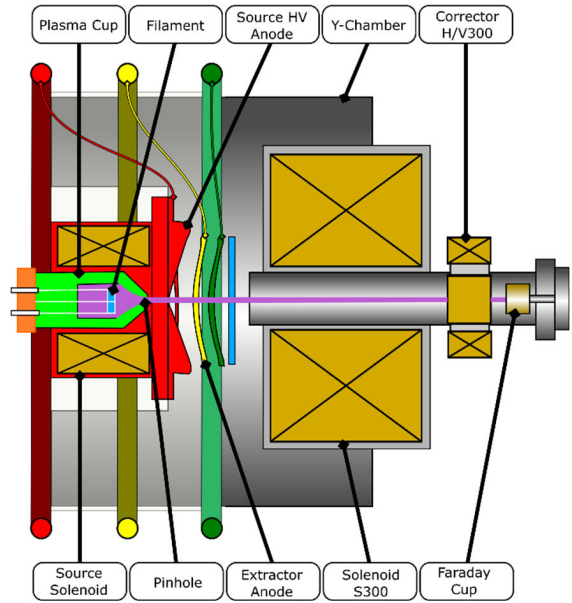

Figure 1: A schematic view of the IS. For extraction of 50 keV protons, the Source High Voltage Anode is charged to a static +50 kV, the Extractor Anode is pulsed to -10 kV relative to the source high voltage (40 kV), and the green anode is at ground potential. The S300 Solenoid is the first element of the LEBT.

## LEBT

The IPI LEBT (Fig. 5) is a short section of beamline designed to transport beam from the IS to the RFQ. Features of note are the two solenoids (SOL) that provide focusing to allow for injection into the RFQ, a gate valve (GV) to isolate the IS from the RFQ, a toroid (TOR) to monitor beam current from the source, and an electrically isolated diaphragm (EID). The EID has two primary roles, first to mask the RFQ injection aperture, but also to measure the intensity of the beam intercepted, allowing for tuning of the solenoids to maximize beam transmission to the RFQ or to perform other beam studies. A pair of beamline corrector packages (H/V) also provide basic horizontal and vertical beam steering capabilities. Finally, scrapers will also be installed for commissioning. These may be removed for operation as the vacuum chamber they occupy restricts the beam aperture.

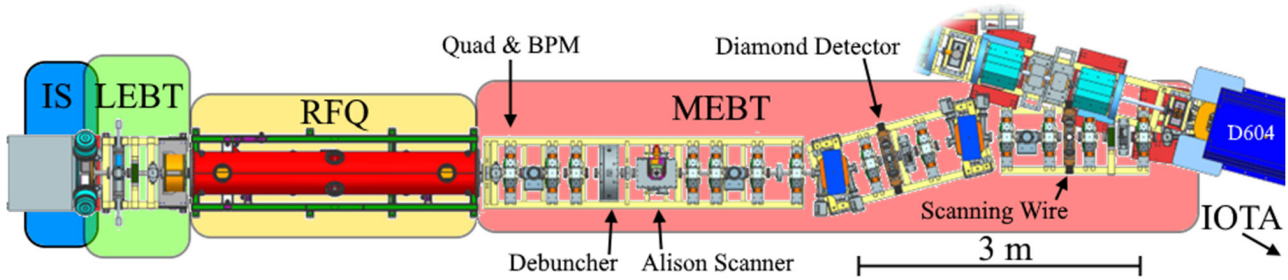

Figure 2: A top-down view of the IPI beamline. It meets the existing Electron Injector beamline at the bend dipole D604 and then injects into the IOTA ring along the same trajectory as electrons at 100 MeV to 150 MeV. As such, time must be taken between electron and proton runs in the IOTA ring to allow for the necessary configuration changes.

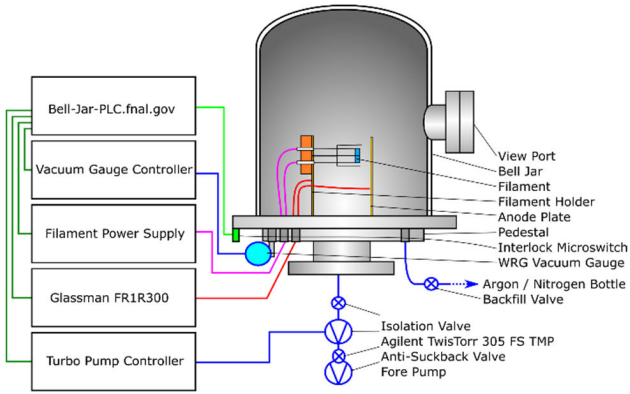

Figure 3: The filament activation station or simply Jar for the large, evacuated bell jar in which activation is performed.

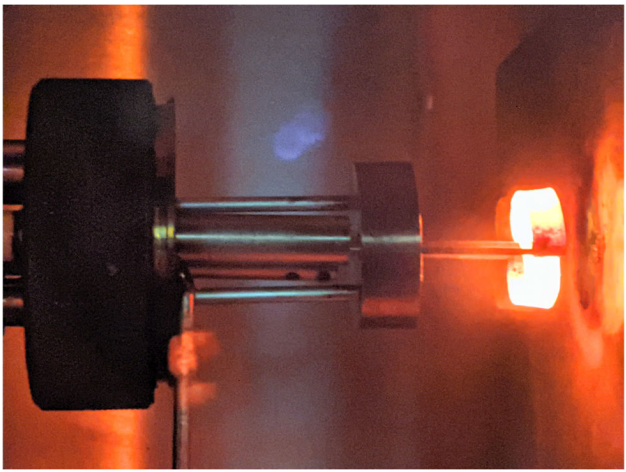

Figure 4: A filament assembly mounted in the Jar for activation.

| | Parameter | Nom. | Unit |
|---|---|---|---|
| LEBT | Energy | 50 | kV |
| | Proton Beam Current | 20 | mA |
| | Pulse length (99%) | 350 | $\mu s$ |
| | Source Pulse Rate | 1 | Hz |
| | Transverse Beam Size | 700 | $\mu m$ |
| MEBT | Energy | 2.5 | MeV |
| | RF Pulse Rate | 1 | Hz |
| | RFQ Frequency | 325.0 ± 0.5 | MHz |
| | RFQ Duty Factor | < 0.002 | % |
| | Phase/Amp. Stability | 1° / 1% | |
| | Beam Pulse | 2 | $\mu s$ |
| | Bunch length (1σ) | 0.3 | ns |
| IOTA (Proton) | Proton Beam Energy | 2.5 | MeV |
| | Relativistic β | $2.66 \cdot 10^{-3}$ | |
| | Circumference | 40 | m |
| | Proton RF Frequency | 2.19 | MHz |
| | Revolution Period | 1.83 | $\mu s$ |
| | RF Voltage | 50 | kV |
| | Geometric Emittance | 0.3 | $\mu m$ |
| | Δp/p (RMS) | 0.3 | % |
| | Beam Current | 8 | mA |
| | RMS Beam size β = 10 m | 4.5 | mm |
| | Momentum compaction | 0.07 | |
| | Betatron tune (Qx, Qy) | 5.3 | |

Table 1: Select parameters specified for or expected from the IPI and the IOTA Ring under proton operation.

## RFQ

Like the IS, the RFQ was also adopted from the HINS experiment and is tasked with accelerating the proton beam from 50 keV to 2.5 MeV. Due to a cooling-to-vacuum leak [5], it cannot currently be temperature-regulated. The low duty factor cited in Table 1 will assist in mitigating the effects of frequency shift, and a test performed in 2020 indicates that the net effect will be a ~5.6 kHz drop in the resonant frequency over the first 8 hours for an average CW power of ~100W [6]. It will level out as the cavity approaches equilibrium as shown in Figure 6, and the initial plan will be to manage the frequency shift with the cavity tuner and the LLRF frequency. This was not a workable plan for the same RFQ when it was used as a part of the HINS experiment as the required duty factor was much higher and the LLRF frequency was fixed. This will be complicated to some extent by the presence of a single bunching cavity in the MEBT, and so further studies will need to be performed as a part of beamline commissioning.

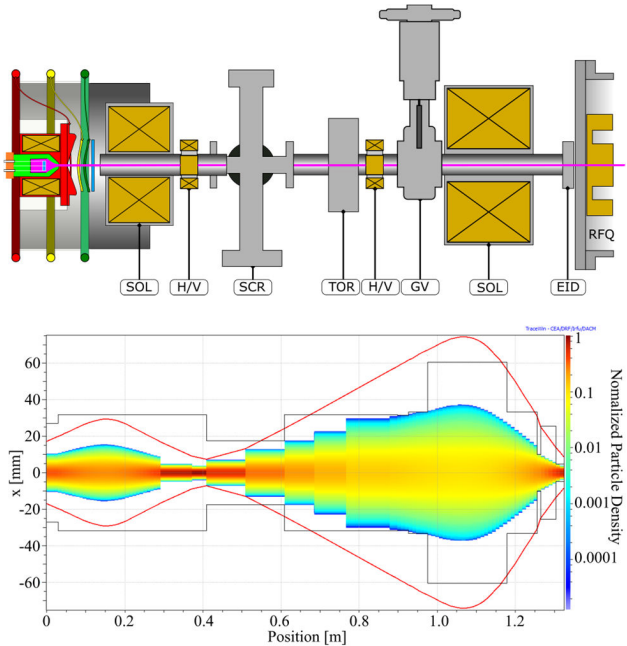

Figure 5: The LEBT configuration and a TraceWin simulation showing aperture restrictions and beam sizes & density through the LEBT.

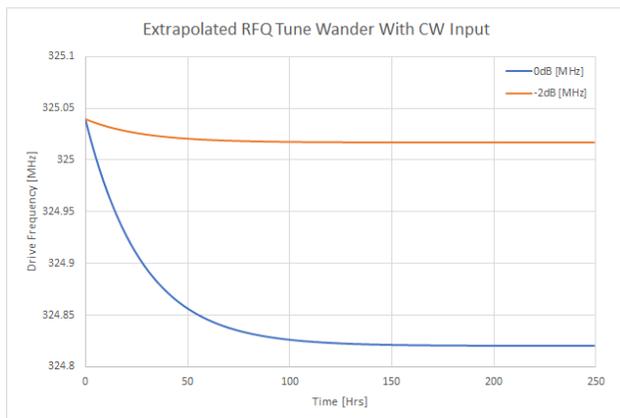

Figure 6: Extrapolated RFQ resonant frequency curves based on 0dB and -2dB set points on an IFR 2042 frequency generator through an Ophir 5083 200W RF amplifier. This corresponds to 149 W and 94 W average CW power respectively in the RFQ.

## MEBT

The IPI MEBT transports the proton beam from the RFQ to the D604 bend dipole, where it joins the existing 300 MeV Electron Injector beamline. It includes the aforementioned bunching RF cavity and a 15° dogleg to assist in matching the beam into the IOTA ring. There are also several pieces of instrumentation for monitoring the beam, including 5 BPMs, each nested into the profile of a beamline quadrupole, an MEBT Alison Scanner for measuring the emittance of the beam following the RFQ, and two scanning wire assemblies to measure transverse beam profiles. A diamond detector is slated for installation in the dogleg section of the beamline.

Beamline magnets include trim dipole packages, dipole bend magnets, and quadrupole magnets repurposed from the low energy Electron Injector beamline build-out. There are also two multi-function magnets, in which each of four coils are independently controlled to affect trim dipole and skew-quadrupole fields in a single element. These multi-function magnets are also used in the IOTA Ring.

## CURRENT STATUS

All instrumentation and beamline elements are in-hand. Acquisition of stands, cables, and other infrastructure is well underway with the goal being to have all materials in-hand by summer 2023. The reassembled source has been charged up to 50 kV, and commissioning and characterization are pending coordination with radiation safety personnel and completion of all Operational Readiness Clearances (ORCs). Installation of the IPI is expected to follow the current electron run in IOTA this summer (2023).

## CONCLUSION

Significant work remains to install and commission the IPI beamline, but success will provide the opportunity of investigating numerous novel accelerator technologies and concepts in the IOTA ring using 2.5 MeV protons.

## ACKNOWLEDGEMENTS

The authors of this paper would like to acknowledge and thank all those who have worked so hard to make the IPI, the IOTA ring, and the Electron Injector beamline a reality, both within the FAST facility departments as well in the various support departments across Fermilab.